\documentstyle[twoside,fleqn,espcrc2,epsf,draft]{article}

\def\WOPPER/{{\tt WOPPER}}
\def\GeV{\mathop{\rm GeV}\nolimits}

\def\PS{[\mathop{\rm PS}]}
\def\PSp{[\mathop{\rm PS'}]}

\makeatletter\let\makeadmark\@makeadmark\makeatother

\title{%
  \WOPPER/, Version 1.1: A Monte Carlo Event Generator for Four
  Fermion Production at LEP-II and Beyond%
  \thanks{Supported by BMfFT, Germany.}%
  \thanks{Presented by Thorsten Ohl.}}

\author{%
  Harald Anlauf\address{%
      SLAC, Theoretical Physics, Mail Stop 81, P.O.~Box 4349,
      Stanford, CA 94309}%
    \thanks{Supported by DFG, Germany.}%
    \thanks{email: {\tt anlauf@slac.stanford.edu}},
  Hans D. Dahmen\address{%
      Universit\"at Siegen,
      Adolph-Reichwein-Str.,
      D-57076 Siegen,
      Germany},
  Angelika Himmler\address{%
      Technische Hochschule Darmstadt,
      Schlo\ss{}gartenstr.~9,
      D-64289 Darmstadt,
      Germany}%
    \xdef\Darmstadt{\theaddress}%
    \thanks{email: {\tt himmler@crunch.ikp.physik.th-darmstadt.de}},
  Panagiotis Manakos\makeadmark{\Darmstadt},
  Thomas Mannel\address{%
      Theory Division, CERN, CH-1211 Geneva 23, Switzerland}%
    \thanks{email: {\tt Thomas.Mannel@Physik.TH-Darmstadt.de}} and
  Thorsten Ohl\makeadmark{\Darmstadt}%
    \thanks{email: {\tt Thorsten.Ohl@Physik.TH-Darmstadt.de}}}

\begin{document}

{\def\$#1: #2 ${#2}\xdef\RCSId{
 \$Id: teupitz94.tex,v 1.13 1994/06/02 22:02:09 ohl Exp $}}

\begin{abstract}
  We report on the status of the Monte Carlo event generator \WOPPER/.
  Version 1.1 of \WOPPER/ describes four fermion production at LEP-II
  and  beyond with leading logarithmic radiative corrections in the
  double~$W^\pm$ pole approximation.  These approximations are
  appropriate for almost all practical purposes, but the inclusion of
  these finite width effects and radiative corrections is nevertheless
  indispensable for LEP-II physics.
\end{abstract}

\maketitle

\section{Introduction}
\label{sec:intr}

The Monte Carlo event generator \WOPPER/~\cite{wopper-cpc} for four
fermion production through~$W^\pm$ resonances at LEP-II and beyond is
the latest addition to the Darmstadt/Siegen family of Monte Carlo
event generators~\cite{unibab/kronos}.  It is a true event generator
that generates a sample of unweighted events which can be used
directly in detector simulations for experiments at LEP-II and future
linear~$e^+e^-$ colliders~\cite{EE500}.  The distinguishing features
of \WOPPER/ are: off-shell pair production in the double pole
approximation and resummation of the leading logarithmic (initial
state) radiative corrections.

\section{Requirements}
\label{sec:requirements}

While we still need to add a direct observation of the triple gauge
vertices (TGV) to the overwhelming indirect evidence for their
existence, any potential deviation from the standard model values will
in all likelihood be very small~\cite{anomalous}.  Thus an observation of
these small anomalous couplings in the permille range will only be
possible at
an high luminosity collider beyond LEP-II, if all standard model
corrections are know to an even better precision. This is only
possible if several independent semi-analytical and Monte Carlo
programs are available, which include (and agree on) all important
contributions.

In the nearer future, the prime physics objective of LEP-II will be a
precise measurement of the $W^\pm$ mass, which will provide an
important cross check of the standard model and help to constrain
possible physics beyond.  For example, reducing the error below
100MeV could close the window for light Higgses (and therefore the
minimal supersymmetric standard model) using constraints from the
electroweak radiative correction parameter~$\Delta r$, as reported at
this conference~\cite{Delta-r}.
Proper accounting for finite width effects and radiative corrections
is of crucial importance for this measurement.  This calls for reliable
calculations and Monte Carlo event generators.

\section{Features of \WOPPER/}
\label{sec:features}

\subsection{Radiative Corrections}
\label{sec:rc}

\WOPPER/ concentrates on the gauge invariant subset of radiative
corrections which is phenomenologically most important.
The leading logarithmic ($\alpha^n\ln^n(s/m_e^2)$)
radiative corrections from the initial state leptons are resummed to
all orders, including the exponentiation of soft photons.

Because of the $t$-channel $\nu$-exchange diagram, it is not possible
to unambiguously separate the complete~${\cal O}(\alpha)$ initial
state radiative corrections in a gauge invariant way.  The restriction
to the leading logarithmic contribution is therefore
phenomenologically reasonable and theoretically sound.

While the resummation of terms beyond ${\cal O}(\alpha^2)$ is
numerically irrelevant for the projected LEP-II statistics, the fully
resummed form can be obtained at no extra cost in a Monte Carlo event
generator and has the desirable side effect of an unproblematic
probabilistic interpretation.  All finite order
approximations suffer from the so called $k_0$ problem: for some
infrared cut $k_0$ on the emitted photon energy, the cross section
\begin{equation}
\label{eq:k0}
  \sigma \approx \left(1 - {\cal O}(1)\cdot\frac{\alpha}{\pi}
     \ln\frac{s}{m_e^2}\ln\frac{E_{\rm Beam}}{k_0}\right)\cdot\sigma_0
\end{equation}
in the channel with no emission of a photon becomes negative.  In
the resummed form, however, the big negative term in~(\ref{eq:k0})
exponentiates and leads to a small but positive cross section.
Therefore the generated
event samples are physically meaningful and the corresponding cross
sections do not depend on the soft photon cut introduced by the
experimental resolution.

Besides the theoretical advantage of keeping only the gauge invariant
leading logs, these corrections have the benefit of leading to a
factorized cross section.  Denoting the phase space variables under
consideration collectively by $\PS$, the cross section reads
\begin{eqnarray}
  \lefteqn{\frac{d\sigma_{LLA}}{d\PS} (\PS)}
           \nonumber \\
\label{eq:x-sect}
    & = & \int_0^1 dx_+ dx_- D(x_+; \mu^2) D(x_-; \mu^2) \\
    &   & \mbox{} \times \frac{d\sigma_0}{d\PSp} (\PSp)\;
          \frac{\partial\PSp}{\partial\PS}
           \nonumber
\end{eqnarray}
where the $\PSp$ depend on $\PS$, and the energy fractions $x^\pm$
of the initial state leptons.  For example, we have $s'=x^+x^-s$ for
the center of mass energy squared.  This cross section
can be implemented easily in an event generator. Here the
electron distribution functions $D(x;\mu^2)$ obey the DGLAP
renormalization group equation:
\begin{eqnarray}
\label{eq:DGLAP}
   \lefteqn{\mu^2 \frac{\partial D(x;\mu^2)}{\partial \mu^2}} \\
      & = & \frac{\alpha}{2\pi}
             \int\limits_x^1 \frac{dz}{z} \left[\frac{1+z^2}{1-z}\right]_+
                 D\left(\frac{x}{z};\mu^2\right)
           \nonumber
\end{eqnarray}

\begin{figure}
  \begin{center}
    \epsfxsize=70mm
    \leavevmode
    \epsffile{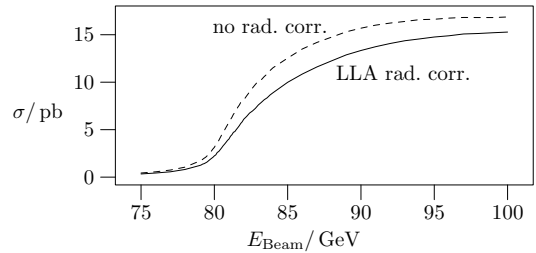}
    \vspace*{-10mm}
  \end{center}
  \caption{Effect of radiative corrections on the total cross section
    near threshold.}
  \label{fig:rc}
\end{figure}

{}From the cross section~(\ref{eq:x-sect}) it can be seen that the
important effects are either universal or of a simple kinematical
origin: the shape of the total cross section will be shifted towards
higher energies because of the energy loss from radiated photons
(``radiative tail'', cf.~figure~\ref{fig:rc}).  Particles which would
be back-to-back
due to momentum conservation in non-radiative events will be
acollinear in radiative events.

The Monte Carlo implementation of~(\ref{eq:x-sect}) in \WOPPER/ first
solves a suitably infrared regularized version of~(\ref{eq:DGLAP}) with a
universal photon shower Monte Carlo and use the sample corresponding
to this solution to define an effective center of mass system (CMS)
for each event after radiation of initial state photons.  In this new
CMS a relatively simple Born type $e^+e^-\to4f$ event can be
generated, which will
then be boosted back to the laboratory frame.  This final boost
incorporates all kinematical effects, like acollinearities for the
intermediate $W$'s, etc.

While this approach is more than adequate for all inclusive
distributions, there remain two areas where further progress is
needed: large photonic $p_T^\gamma$ and final state radiation.

\begin{figure}
  \begin{center}
    \epsfxsize=73.5mm
    \leavevmode
    \epsffile{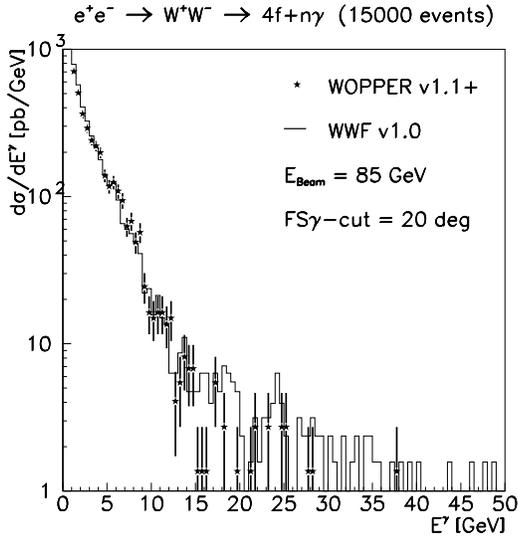}
    \vspace*{-10mm}
  \end{center}
  \caption{$E^\gamma$ spectrum at $\protect\sqrt{s} = 170\GeV$,
    comparing \WOPPER/'s LLA prediction with {\tt wwf}'s
    ${\cal O}(\alpha)$ prediction. An angular cut for photons around
    any charged particle in the final state has been applied.}
  \label{fig:e85}
\end{figure}

\begin{figure}
  \begin{center}
    \epsfxsize=73.5mm
    \leavevmode
    \epsffile{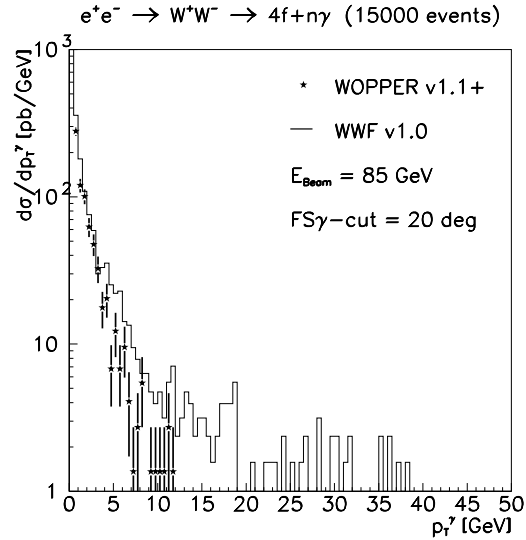}
    \vspace*{-10mm}
  \end{center}
  \caption{Same as figure~\protect\ref{fig:e85}, but for the photonic
     transversal momentum $p_T^\gamma$.}
  \label{fig:pt85}
\end{figure}

\begin{figure}
  \begin{center}
    \epsfxsize=73.5mm
    \leavevmode
    \epsffile{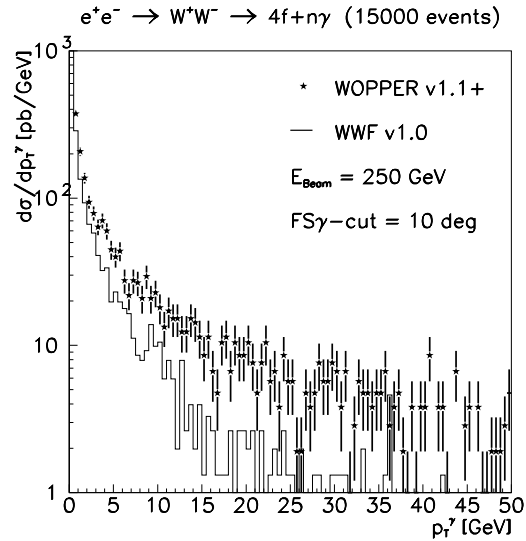}
    \vspace*{-10mm}
  \end{center}
  \caption{Same as figure~\protect\ref{fig:pt85}, but at higher
    energy $\protect\sqrt{s} = 500\GeV$.}
  \label{fig:pt250}
\end{figure}

By its very definition, the leading logarithmic or pole approximation
is applicable to collinear radiation and inclusive spectra because
these are dominated by collinear radiation
(cf.~fig.~\ref{fig:e85}, where the LLA is compared to a ${\cal
O}(\alpha)$ calculation).  But for large photonic $p_T^\gamma$ the LLA
is essentially an uncontrolled approximation.  However, we can use
existing ${\cal O}(\alpha)$ Monte Carlos~\cite{wwf} to gauge the
numerical accuracy of the LLA even for large $p_T^\gamma$.  It turns
out that the pole approximation reproduces the full ${\cal O}(\alpha)$
calculation surprisingly well in the projected LEP-II energy regime
(cf.~fig.~\ref{fig:pt85}).
At 500GeV on the other hand, the large $p_T^\gamma$ region is
overestimated (cf.~fig.~\ref{fig:pt250}).

The other problematic issue concerns final state radiation.  While the
figures \ref{fig:e85}, \ref{fig:pt85} and \ref{fig:pt250} confirm that
a reasonable angular cut around any charged particle in the final
state will remove all effects of final state radiation, are more
complete Monte Carlo treatment is of course desirable.

\subsection{Finite Width}
\label{sec:fin-width}

For detailed background studies, it is of course necessary to study
all non-resonant diagrams contributing to $e^+e^-\to4f$ at LEP-II
energies~\cite{excalibur}.  For the identification of $W^+W^-$ pairs,
experimental invariant mass cuts will have to be applied, however.
After such cuts, the contribution of non-resonant diagrams goes down
rapidly~\cite{excalibur}.

For the sake of efficiency, version 1.1 of \WOPPER/ implements the
four fermion final states therefore in double pole approximation.
This is equivalent to keeping the two so-called ``signal'' diagrams.

\begin{figure}
  \begin{center}
    \epsfxsize=70mm
    \leavevmode
    \epsffile{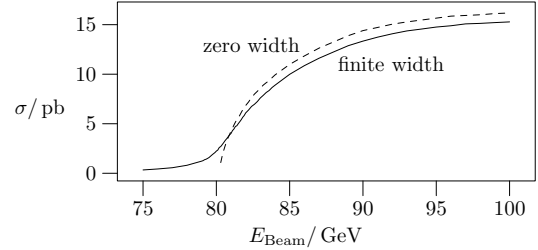}
    \vspace*{-10mm}
  \end{center}
  \caption{Effect of the finite $W^\pm$ width on the total cross section
    near threshold.}
  \label{fig:fw}
\end{figure}

For all practical purposes, the $W^\pm$'s decay into light quarks only
($b$'s are still relatively light and because of the small
$|V_{cb}|^2$ rare).  Thus the unphysical polarizations decouple and the
cross section factorizes
\begin{eqnarray}
  \lefteqn{\sigma_{\rm\scriptsize resonant}} \nonumber \\
    & = & \int ds_+ ds_-
      \frac{\sqrt{s_+} \,\Gamma_W(s_+)}{\pi D(s_+)}
      \frac{\sqrt{s_-} \,\Gamma_W(s_-)}{\pi D(s_-)} \\
    & & \mbox{} \times  \sigma_{\rm\scriptsize off-shell}(s; s_+, s_-)
        \nonumber
\end{eqnarray}
into ``off-shell'' production $\sigma_{\rm\scriptsize off-shell}(s;
s_+, s_-)$, Breit-Wigner propagators
\begin{equation}
  \frac{1}{D(s_\pm)} =
  \frac{1}{(s_\pm - M_W^2)^2 + s_\pm \Gamma_W^2(s_\pm)} \; .
\end{equation}
and decay widths $\Gamma_W(s_\pm)$.  This factorized form is again
very convenient for implementation in a Monte Carlo event generator.
The resulting cross section, displaying the typical smearing of the
threshold, is depicted in figure~\ref{fig:fw}.

\subsection{Parton Showers, Fragmentation and Hadronization}
\label{sec:QCD}

The description of semileptonic and hadronic $W^+W^-$ events is of
course incomplete without proper accounting for the hadronization of
the quarks in the final state.  Working interfaces of \WOPPER/ to both
major final state parton shower and fragmentation models {\tt
JETSET~7.4}~\cite{JETSET} and {\tt HERWIG~5.5}~\cite{HERWIG} are
implemented as of version 1.1.  Therefore the events generated by
\WOPPER/ can be fed immediately to detector simulation Monte Carlos.

\section{Sample Application}
\label{sec:sample}

At high energies, the transversally polarized $W^\pm$'s from the
$t$-channel diagram are dominating the cross section.  From a physics
perspective however, the longitudinally polarized $W^\pm$'s and the
contribution from the $s$-channel diagram are both far more
interesting.  The former because the longitudinal $W^\pm$'s are a
direct manifestation of electroweak symmetry breaking and the latter
is where the three gauge boson vertex is to be measured.

It is therefore important to be able to reconstruct a sample of
longitudinally polarized $W^\pm$'s.   Fortunately, the $V-A$ decay of
$W^\pm \to f\bar f'$  is self analyzing, if we can measure the angular
{\em decay\/} distribution:
\begin{equation}
 \sigma_{+,0,-} = \int d\cos\theta^* \; P_{+,0,-} (\cos\theta^*)
    \frac{d\sigma}{d\cos\theta^*}
\end{equation}
Here the $P_{+,0,-}$ are simple polynomials~\cite{Davier}.

\begin{figure}
  \begin{center}
    \epsfxsize=73.5mm
    \leavevmode
    \epsffile{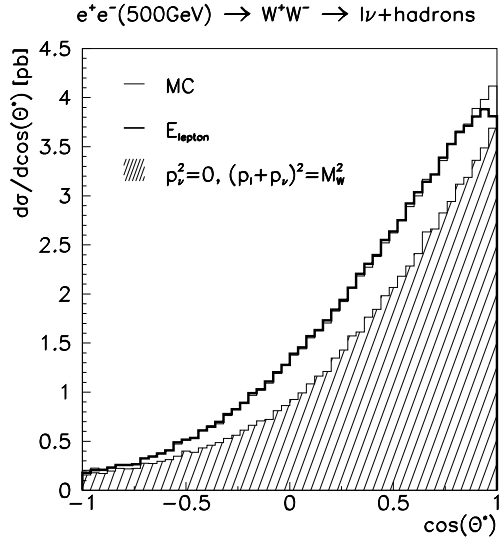}
    \vspace*{-10mm}
  \end{center}
  \caption{Different methods of reconstructing the decay angle
    $\theta^*$ in the presence of acceptance cuts
    ($175^\circ>\theta>5^\circ$) and finite $W^\pm$ width.}
  \label{fig:cts0}
\end{figure}

\begin{figure}
  \begin{center}
    \epsfxsize=73.5mm
    \leavevmode
    \epsffile{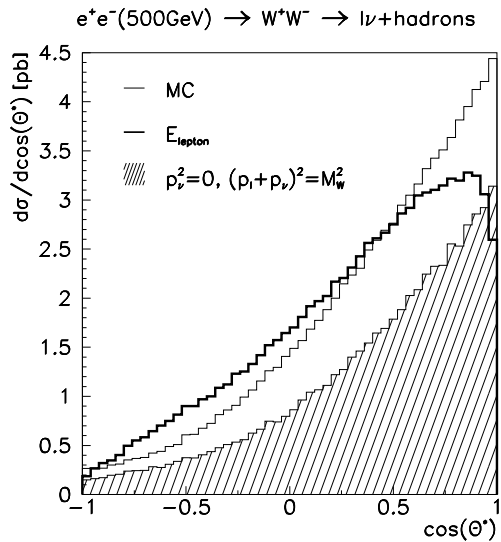}
    \vspace*{-10mm}
  \end{center}
  \caption{Same as figure~\protect\ref{fig:cts0}, but additionally
    including leading log initial state radiative corrections.}
  \label{fig:cts}
\end{figure}

In the case of on-shell $W^\pm$'s without radiative corrections, the
decay angle can be reconstructed in semileptonic decays by measuring
the energy $E_\ell$ of the charged lepton only~\cite{Davier}
\begin{equation}
  \cos\theta^* = \frac{2E_\ell - E_B}{\sqrt{E_B^2- M^2_W}}.
\end{equation}
However, using semi realistic acceptance cuts of
$175^\circ>\theta>5^\circ$, this method fails in the presence of
finite $W^\pm$ width and radiative corrections, as can be seen from
the fat histograms in figures~\ref{fig:cts0} and~\ref{fig:cts}.

On the other hand, it seems to be more promising to try to reconstruct
the neutrino momentum from the missing transversal momentum, the
mass shell condition $p_\nu^2=0$ of the neutrino and the approximate
mass shell condition of the leptonically decaying~$W$:
$(p_\ell+p_{\nu_\ell})^2=M_W^2$.  If we add the
further requirement that the squared sum of all reconstructed momenta
should deviate no more than 10\%\ from the squared CMS energy, we get
the much improved hashed histograms in figures~\ref{fig:cts0}
and~\ref{fig:cts}.  They are in particular free from the
characteristic shape distortions generated by the first method.

\section{Conclusions and outlook}
\label{sec:concl}

\WOPPER/ is a fast, flexible and supported tool for~$W^\pm$ physics at
LEP-II and beyond.  Comparisons with other Monte Carlos~\cite{wwf} has
shown that the approximations used in \WOPPER/ (leading log radiation,
resonant~$W$'s) can easily be controlled for experimentally
relevant cuts.
In the forthcoming releases, the following features will be added:
\begin{itemize}
  \item{} Anomalous couplings.  This feature has been requested
    by experimentalists during this workshop again.  Therefore this
    straightforward, if somewhat tedious, enhancement will be
    installed.
  \item{} Non electromagnetic radiative corrections.  We will add the
    dominant contributions which go beyond the running QED coupling.
    The latter is of course already available.
  \item{} Coulomb singularity. We will add this feature soon, again by
    popular demand\footnote{At the time this contribution to the
      proceedings is written,
      the Coulomb singularity has already been implemented in \WOPPER/,
      Version 1.2.}.  For the time being, it will implemented
    inclusively, without generation of the corresponding soft photons.
  \item{} Improved photonic~$p_T^\gamma$ spectrum.  Comparison with
    complete~${\cal O}(\alpha)$ Monte Carlos shows that the LLA is an
    excellent approximation for inclusive distributions and
    longitudinal spectra.  There remains however a noticeable
    discrepancy at high photonic~$p_T^\gamma$, which can not be
    described in
    LLA.  This situation will be improved, either by explicit
    inclusion of non leading terms or by phenomenological
    interpolation.
  \item{} Final state radiation and interference terms.  This will
    come somewhere further down the road.
\end{itemize}

Due to the slowly varying nature of the~$e^+e^- \to W^+W^-$ cross
section, the forward branching algorithm~\cite{unibab/kronos} with
hand crafted importance sampling as implemented in \WOPPER/ v1.x is
sufficient for most practical purposes.  We shall however replace it
by a backward branching algorithm, which will blend better with a
general purpose Monte Carlo engine, where the cross sections can be
replaced more easily~\cite{clov}.

\section{Distribution policy}

\begingroup
  \def\|{\allowbreak/\allowbreak}
  \def\.{\allowbreak.\allowbreak}
  \def\:{\allowbreak:\allowbreak}
  \def\@{\allowbreak@\allowbreak}
  \def\-{\allowbreak-}

\WOPPER/ is distributed electronically over the Internet using the
following channels:
\begin{itemize}
  \item{} The {\tt FORTRAN-77} sources (in {\tt PATCHY} format) can be
     obtained by anonymous Internet ftp from
     the host {\tt crunch\.ikp\.physik\.th\-darmstadt\.de} in the
     directories {\tt pub\|ohl\|wopper\|old}, {\tt
     pub\|ohl\|wopper\|pro} and {\tt pub\|ohl\|wopper\|new},
     corresponding to slightly outdated, current and experimental
     releases of \WOPPER/ respectively.
  \item{} The current status of \WOPPER/ can be queried
    through the World Wide Web from the document {\tt
    http\:\|\|crunch\.ikp\.physik\.th\-darmstadt\.de\|monte\-carlos\.html}.
  \item{} Important announcements (new versions, fatal bugs, etc.)
    will be made through the mailing list {\tt
    wopper\-announce\@crunch\.ikp\.physik\.th\-darmstadt\.de}.
    Subscriptions should be mailed to {\tt wopper\-announce\-request} at
    the same host. The purpose of {\tt wopper\-announce} is not general
    discussions of \WOPPER/, however, if there is interest among
    users, a companion list {\tt wopper\-discuss} can be created by the
    authors easily.
\end{itemize}

\endgroup


\end{document}